\documentclass{ws-ijmpa}
\usepackage[super,compress]{cite}
\usepackage{graphicx}

\def\Journal#1#2#3#4{{#1} {\bf #2}, #3 (#4)}
\def\AA{Astron. Astrophys. }

\def\CPC{Chin. Phys. C}
\def\EPJC{Eur. Phys. J. C}
\def\IJMPA{Int. J. Mod. Phys. A}

\def\JCAP{J. Cosmol. Astropart. Phys.}

\def\JCAP{JCAP}
\def\JHEP{JHEP}

\def\JPG{J. Phys. G} 
 
\def\MPLA{Mod. Phys. Lett. A}

\def\NPB{Nucl. Phys. B}

\def\PLB{Phys. Lett. B}
\def\PREP{Phys. Rep.}

\def\PAN{Phys. Atom. Nucl.}
\def\PRL{Phys. Rev. Lett.}
\def\PRD{Phys. Rev. D}

\begin{document}
\markboth{Teruyuki Kitabayashi, Shinya Ohkawa and Masaki Yasu\`{e}}{One-loop radiative seesaw dark matter and neutrinoless double beta decay with two zero flavor neutrino mass texture}

%
\catchline{}{}{}{}{}
%

\title{One-loop radiative seesaw dark matter and neutrinoless double beta decay with two zero flavor neutrino mass texture}

\author{Teruyuki Kitabayashi, Shinya Ohkawa and Masaki Yasu\`{e}}

\address{Department of Physics, Tokai University,\\
4-1-1 Kitakaname, Hiratsuka, Kanagawa 259-1292, Japan\\
teruyuki@tokai-u.jp, 6BSNM004@mail.u-tokai.ac.jp, yasue@keyaki.cc.u-tokai.ac.jp}

\maketitle

\begin{history}
\received{Day Month Year}
\revised{Day Month Year}
\end{history}

\begin{abstract}
We discuss the linkage between dark matter mass in the one-loop radiative seesaw model and the effective neutrino mass for the neutrino less double beta decay. This linkage, which has been already numerically suggested, is confirmed to be a reasonable relationship by deriving analytical expressions for two zero flavor neutrino mass texture.
\end{abstract}

\ccode{PACS numbers:14.60.Pq, 98.80.Cq}



\section{Introduction}
\label{sec:introduction}
Understanding the nature of dark matter as well as of neutrinos is one of the outstanding problems in particle physics. Recently, Ma has been proposed a simple model, so-called radiative seesaw model or scotogenic model, which can simultaneously account for the origin of neutrino masses and the presence of dark matter \cite{Ma2006PRD}. In this model, neutrino masses vanish at the tree level but are generated by one-loop interactions mediated by a dark matter candidate. One-loop \cite{one_loop_Ma1998PRL,one_loop_Kubo2006PLB,one_loop_Hambye2007PRD,one_loop_Farzan2009PRD,one_loop_Farzan2010MPLA,one_loop_Farzan2011IJMPA,one_loop_Kanemura2011PRD,one_loop_Schmidt2012PRD,one_loop_Faezan2012PRD,one_loop_Aoki2012PRD,one_loop_Hehn2012PLB,one_loop_Bhupal2012PRD,one_loop_Bhupal2013PRD,one_loop_Law2013JHEP,one_loop_Kanemura2013PLB,one_loop_Hirsch2013JHEP,one_loop_Restrepo2013JHEP,one_loop_Lindner2014PRD,one_loop_Okada2014PRD89,one_loop_Okada2014PRD90,one_loop_Brdar2014PLB,one_loop_Borah2015PRD,one_loop_Wang2015PRD,one_loop_Fraser2016PRD,one_loop_Adhikari2016PLB,one_loop_Ma2016PLB,one_loop_Arhrib2016JCAP,one_loop_Okada2016PRD,one_loop_Ahriche2016PLB,one_loop_Lu2016JCAP,one_loop_Cai2016JHEP} as well as two-loops \cite{two_loop_Ma2007PLB,two_loop_Ma2008PLB,two_loop_Lindner2011PLB,two_loop_Kanemura2012PRD,two_loop_Kajiyama2013NPB,two_loop_Aoki2013PRD,two_loop_Kajiyama2013PRD,two_loop_Baek2014PLB,two_loop_Baek2014JCAP,two_loop_Kanemura2014PRD,two_loop_Aoki2014JCAP,two_loop_Aoki2014PRD,two_loop_Farzan2015JHEP,two_loop_Okada2016PRD} and three-loops \cite{three_loop_Krauss2003PRD,three_loop_Aoki2009PRL,three_loop_Gustafsson2013PRL,three_loop_Ahriche2014PRD,three_loop_Ahriche2014JHEP,three_loop_Nishiwaki2015PRD,three_loop_Culjak2015PLB,three_loop_Antipin2017PLB} interactions related to neutrino mass and dark matter have been extensively studied in literature.

On the other hand, there have been various discussions on neutrino masses to ensure the appearance of the observed neutrino mixings and masses, for example, based on flavor neutrino mass matrices with two zeros \cite{twoZeroFlavor_Fritzsch2011JHEP,twoZeroFlavor_Zhou2016CPC,Dev2014PRD,Meloni2014PRD}. This type of matrix is called the two zero flavor neutrino mass texture. If we require a nonvanishing effective neutrino mass $M_{ee}$ for the neutrino less double beta decay \cite{doubleBetaDecay_Cremonesi2014AHEP,doubleBetaDecay_Benato2015EPJC}, only four textures are compatible with observed data in the two zero flavor neutrino mass texture scheme.

In this paper, we clarify the linkage between dark matter mass in the one-loop radiative seesaw model and the effective neutrino mass for the neutrino less double beta decay. This linkage has been numerically suggested by Kubo, Ma and Suematsu \cite{one_loop_Kubo2006PLB}. Using the two zero flavor neutrino mass texture, we show this connection more explicitly by deriving analytical expressions supplemented by additional numerical calculations.

In Sec.\ref{sec:review}, we show a brief review of the radiative seesaw model and the two zero flavor neutrino mass texture. In Sec.\ref{sec:Dark_matter}, we show the linkage between dark matter mass and the effective neutrino mass for the neutrino less double beta decay. Sec.\ref{sec:summary} is devoted to summary.

\section{\label{sec:review} Brief review}
\subsection{Radiative seesaw model}
The radiative seesaw model \cite{Ma2006PRD} is an extension of the standard model containing three new Majorana $SU(2)_L$ singlet fermions $N_k$ $(k=1,2,3)$ and one new scalar $SU(2)_L$ doublet $(\eta^+,\eta^0)$. These new particles are odd under exact $Z_2$ symmetry. Under $SU(2)_L \times U(1)_Y \times Z_2$, the main particle contents for radiative seesaw model is given by  $(\alpha=e,\mu,\tau; k=1,2,3)$ :
\begin{eqnarray}
&& L_\alpha=(\nu_\alpha, \ell_\alpha) \ : \ (2,-1/2,+), \quad \ell_\alpha^C \ : \ (1,1,+),\nonumber \\
&& \Phi=(\phi^+, \phi^0) \ : \ (2,1/2,+), \nonumber \\
&& N_k \ : \  (1,0,-), \quad \eta=(\eta^+,\eta^0) \ : \ (2,1/2,-),  
\end{eqnarray}
where $(\nu_\alpha, \ell_\alpha)$ is the left-handed lepton doublet and $(\phi^+, \phi^0)$ is the Higgs doublet in the standard model. 

The new particles are contained in Yukawa interactions
\begin{eqnarray}
\mathcal{L}_{Y} \supset h_{\alpha k} (\nu_\alpha \eta^0 - \ell_\alpha \eta^+) N_k + h.c.,
\end{eqnarray}
in the Majorana mass terms
\begin{eqnarray}
\mathcal{L}_{N} \supset \frac{1}{2}M_k N_k N_k + h.c.,
\end{eqnarray}
and in the quartic scalar interaction potential
\begin{eqnarray}
V_{int}  \supset \frac{1}{2}\lambda_5 (\Phi^\dagger \eta)^2 + h.c.
\end{eqnarray}

Owing to the $Z_2$ symmetry, neutrinos remain massless at tree level but acquire masses via one-loop interactions. The neutrino flavor masses read \cite{Ma2006PRD,Merle2015arXiv,Vicente2017arXiv}
\begin{eqnarray}
M_{\alpha\beta} = \sum_{k=1}^3 h_{\alpha k}h_{\beta k} \Lambda_k,
\label{Eq:M_alpha_beta}
\end{eqnarray}
with
\begin{eqnarray}
\Lambda_k =  \frac{\lambda_5 v^2}{16\pi^2}\frac{M_k}{m^2_0-M^2_k}\left(1-\frac{M^2_k}{m^2_0-M^2_k}\ln\frac{m_0^2}{M^2_k} \right),
\label{Eq:Lambda_k}
\end{eqnarray}
where we define $m_0^2 = (m_R^2+m_I^2)/2$ and $v$, $m_R$, $m_I$ denote vacuum expectation value of the Higgs field, the masses of $\sqrt{2} {\rm Re}[\eta^0]$ and $\sqrt{2} {\rm Im}[\eta^0]$, respectively. 

At the one-loop level, flavor violating processes such as $\mu \rightarrow e \gamma$ are induced. The branching ratio of $\mu \rightarrow e \gamma$ is given by \cite{Ma2001PRL,one_loop_Kubo2006PLB}

\begin{eqnarray}
{\rm Br}(\mu \rightarrow e \gamma)=\frac{3\alpha_{\rm em}}{64\pi(G_Fm_0^2)^2}\left| \sum_{k=1}^3 h_{\mu k}h_{e k}^* F \left( \frac{M_k^2}{m_0^2}\right) \right|^2,
\end{eqnarray}
where $\alpha_{\rm em}$ denotes the fine-structure constant (electromagnetic coupling), $G_F$ denotes the Fermi coupling constant and  $F(x)$ is defined by
\begin{eqnarray}
F(x)=\frac{1-6x+3x^2+2x^3-6x^2 \ln x}{6(1-x)^4}.
\end{eqnarray}
%

\subsection{Relic abundance of dark matter}
The radiative seesaw mechanism of neutrino masses also predicts the existence of particle dark matter. The $Z_2$ symmetry renders the lightest $Z_2$ odd particle stable in the particle spectrum and this lightest $Z_2$ odd particle becomes a dark matter candidate. 

We assume that the lightest Majorana singlet fermion, which is taken to be $N_1$, becomes the dark matter. We consider the following two cases:
\begin{description}
\item[(i)] $M_1 \ll M_2 < M_3$, 
\item[(ii)] $M_1 \lesssim M_2 < M_3$.
\end{description}
Especially in the case (ii), we have to take account of coannihilation effects \cite{Griest1991PRD} because $N_1$ is considered to be almost degenerate with the next to lightest Majorana singlet fermion $N_2$. The effective cross section $\sigma_{\rm eff}$, including contributions from coannihilation, is obtained as
\begin{eqnarray}
\sigma_{\rm eff} &=& \frac{g_{N_1}^2}{g_{\rm eff}^2}\sigma_{N_1N_1} + \frac{2g_{N_1}g_{N_2}}{g_{\rm eff}^2}\sigma_{N_1N_2} (1+\Delta M)^{3/2}e^{-\Delta M \cdot x} \nonumber \\
&&+ \frac{g_{N_2}^2}{g_{\rm eff}^2}\sigma_{N_2N_2} (1+\Delta M)^3 e^{-2\Delta M \cdot x},\nonumber \\
g_{\rm eff}&=&g_{N_1}+g_{N_2} (1+\Delta M)^{3/2}e^{-\Delta M \cdot x}, 
\end{eqnarray}
where $\sigma_{N_iN_j}$ $(i,j=1,2)$ is annihilation cross section for $N_i N_j \rightarrow \bar{f}f$, $\Delta M = (M_2-M_1)/M_1$ depicts the mass splitting ratio of the degenerate singlet fermions, $x = M_1/T$ denotes the ratio of the mass of lightest singlet fermion to the temperature $T$ and $g_{N_1}$ and $g_{N_2}$ are the number of degrees of freedom of $N_1$ and $N_2$, respectively. 

The effective (co)annihilation cross section times the relative velocity of annihilation particles $v_{\rm rel}$ is given by \cite{Suematsu2009PRD}
\begin{eqnarray}
\sigma_{N_iN_j} |v_{\rm rel}|&=& \frac{1}{8\pi}\frac{M_1^2}{(M_1^2+m_0^2)^2} \left( 1+\frac{m_0^4-3m_0^2M_1^2-M_1^4}{3(M_1^2+m_0^2)^2} v_{\rm rel}^2\right) \nonumber \\
&& \times \sum_{\alpha\beta}(h_{\alpha i} h_{\beta j} - h_{\alpha j} h_{\beta i})^2 +  \frac{1}{12\pi}\frac{M_1^2(M_1^4+m_0^4)}{(M_1^2+m_0^2)^4} \nonumber \\
&& \times v_{\rm rel}^2\sum_{\alpha\beta}h_{\alpha i} h_{\alpha j} h_{\beta i} h_{\beta j},
\end{eqnarray}
where $i,j$ should be taken to be $1$ or $2$. Since we are interested in the effect of coannihilation, we assume $\Delta M \simeq 0$ and obtain
\begin{eqnarray}
\sigma_{\rm eff} |v_{\rm rel}|&=& \left(\frac{\sigma_{N_1N_1}}{4} + \frac{\sigma_{N_1N_2}}{2} + \frac{\sigma_{N_2N_2}}{4}\right) |v_{\rm rel}|.
\end{eqnarray}
If we define $a_{\rm eff}$ and $b_{\rm eff}$ by $\sigma_{\rm eff} = a_{\rm eff} + b_{\rm eff} v_{\rm rel}^2$, thermally averaged cross section $\langle \sigma_{\rm eff}|v_{\rm rel}| \rangle$ can be written as $\langle \sigma_{\rm eff}|v_{\rm rel}| \rangle = a_{\rm eff} + 6b_{\rm eff}/x$, where 
\begin{eqnarray}
a_{\rm eff}&=& \frac{1}{16\pi}\frac{M_1^2}{(M_1^2+m_0^2)^2}\sum_{\alpha\beta}(h_{\alpha 1} h_{\beta 2} - h_{\alpha 2} h_{\beta 1})^2,  \nonumber \\
b_{\rm eff}&=&  \frac{1}{16\pi}\frac{M_1^2}{(M_1^2+m_0^2)^2}\frac{m_0^4-3m_0^2M_1^2-M_1^4}{3(M_1^2+m_0^2)^2} \sum_{\alpha\beta}(h_{\alpha 1} h_{\beta 2} - h_{\alpha 2} h_{\beta 1})^2 \nonumber \\
&& +  \frac{1}{48\pi}\frac{M_1^2(M_1^4+m_0^4)}{(M_1^2+m_0^2)^4} \sum_{\alpha\beta}(h_{\alpha 1} h_{\beta 1} + h_{\alpha 2} h_{\beta 2})^2.
\end{eqnarray}

The relic abundance of cold dark matter is estimated to be:
\begin{eqnarray}
\Omega h^2 = \frac{107\times 10^9 x_f}{g_\ast^{1/2} m_{\rm pl}({\rm GeV}) (a_{\rm eff}+3b_{\rm eff})/x_f },
\end{eqnarray}
where $m_{\rm pl}=1.22\times 10^{19} {\rm GeV}$, $g_{\ast} = 106.75$ and
\begin{eqnarray}
x_f = \ln \frac{0.038 g_{\rm eff} m_{\rm pl} M_1 \langle \sigma_{\rm eff} |v_{\rm rel}| \rangle}{g_\ast^{1/2} x_f^{1/2} }.
\end{eqnarray}
%

\subsection{Two zero flavor neutrino mass texture}
In the two zero flavor neutrino mass texture scheme, there are 15 possible combinations of two vanishing independent elements in the flavor neutrino mass matrix. If we require a nonvanishing effective neutrino mass $M_{ee}$ for the neutrino less double beta decay, the interesting textures are the following only four \cite{twoZeroFlavor_Fritzsch2011JHEP,twoZeroFlavor_Zhou2016CPC}
\begin{eqnarray}
&& {\rm B}_1:
\left( 
\begin{array}{*{20}{c}}
M_{ee} & M_{e\mu} & 0 \\
- & 0 & M_{\mu\tau} \\
- & - & M_{\tau\tau} \\
\end{array}
\right),
\
{\rm B}_2:
\left( 
\begin{array}{*{20}{c}}
M_{ee} & 0 & M_{e\tau} \\
-& M_{\mu\mu} & M_{\mu\tau} \\
- & - & 0 \\
\end{array}
\right),
\nonumber \\
&&{\rm B}_3:
\left( 
\begin{array}{*{20}{c}}
M_{ee} & 0 & M_{e\tau} \\
- & 0 & M_{\mu\tau} \\
- & - & M_{\tau\tau} \\
\end{array}
\right),
\
{\rm B}_4:
\left( 
\begin{array}{*{20}{c}}
M_{ee} & M_{e\mu} & 0 \\
- & M_{\mu\mu} & M_{\mu\tau} \\
- & - & 0 \\
\end{array}
\right),
\label{Eq:B1B2B3B4}
\end{eqnarray}
where the mark ``$-$" denotes a symmetric partner. For these textures, nearly degenerated neutrino masses are favorable \cite{Dev2014PRD}; however, neutrino mass ordering, either the normal ordering (NO), $m_1 \le m_2 \le m_3$, or the inverted ordering (IO), $m_3 \le m_1 \le m_2$, is not determined. 

The texture zeros in the Majorana neutrino mass matrix can be realized by imposing appropriate symmetries on the Lagrangian of neutrino mass models \cite{Dev2014PRD,textureZerosAndAbelianSymmetries_Grimus2004EPJC,textureZerosAndAbelianSymmetries_Grimus2005JPG,textureZerosAndAbelianSymmetries_Dev2011PLB,textureZerosAndAbelianSymmetries_Felipe2014NPB,textureZerosAndAbelianSymmetries_Lavoura2015JPG,Zhou2016CPC,textureZerosAndStability}. For example, $B_1, B_2, B_3$ and $B_4$ textures are obtained by imposing the cyclic group $Z_3$ \cite{Dev2014PRD}. There are other symmetry realization of $B_1, B_2, B_3$ and $B_4$ textures. The $B_1$ and $B_2$ textures have been realized by using $A_4$ or its $Z_3$ subgroup \cite{Hirsch2007PRL}. The $B_3$ and $B_4$ textures have been obtained by soft breaking of the $L_\mu-L_\tau$ symmetry \cite{Rodejohann2005PAN}. Moreover, one-loop induced radiative neutrino mass model with some flavor dependent $U(1)$ gauge symmetry and two zero textures are also discussed \cite{Baek2015JHEP, Ko2017arXiv}. In this paper, because we concentrate on researching the relation between dark matter mass in the scotogenic model and effective neutrino mass of neutrinoless double beta decay in the two zero textures, we would like to put aside the  discussion of the particular mechanism for realization of texture two zero scheme. Whether the simplicity of the Eq.(\ref{Eq:M_alpha_beta}) is held or not with two zero texture is interesting question. The realization of two zero texture in the scotogenic model is important issue and we will discuss it in our future study.

Our recent discussions \cite{KitabayashiYasue2016IJMPA,KitabayashiYasue2016PRD} have found the dependence of the flavor neutrino masses $M_{\alpha\beta}$ $(\alpha,\beta=e,\mu,\tau)$ and mass eigen values $m_j$ $(j=1,2,3)$ on $M_{ee}$, which dictates, for textures labelled by $X={\rm B_1,B_2,B_3,B_4}$,
\begin{eqnarray}
M_{\alpha\beta} = f_{\alpha\beta}^X(\theta_{12},\theta_{23}, \theta_{13},\delta)M_{ee},
\label{Eq:M_flavor}
\end{eqnarray}
and
\begin{eqnarray}
m_je^{-i\phi_j} &=& f_j^X(\theta_{12},\theta_{23}, \theta_{13},\delta)M_{ee}, \label{Eq:mass_ev}
\end{eqnarray}
with the obvious definition of $f^X_{ee}=1,$ where $\theta_{12, 23, 23}$ are three neutrino mixing angles and $\delta$ is the CP-violating Dirac phase as defined in Ref.\cite{PDG} and $\phi_1,\phi_2, \phi_3$ are the Majorana CP phases.  

More explicitly, we obtain the following expressions for the ${\rm B_1}$ texture:
\begin{eqnarray}
M_{\alpha\beta}&=&f_{\alpha\beta}^{\rm B1}M_{ee}, \nonumber \\ 
m_je^{-i\phi_j} &=& f_j^{\rm B1}M_{ee},
\label{Eq:M_B1} 
\end{eqnarray}
where
\begin{eqnarray}
f_{e\tau}^{\rm B1} &=& f_{\mu\mu}^{\rm B1}=0,
\\
f_{e\mu}^{\rm B1}&=&-\frac{A_1}{c_{23}B_1 + s_{23}C_1},
\nonumber \\
f_{\mu\tau}^{\rm B1}&=&-A_1\frac{-c_{23}B_3 + s_{23}C_3}{c_{23}B_1 + s_{23}C_1} - \frac{1-e^{-2i\delta}}{2} \sin 2\theta_{23},
\nonumber \\
f_{\tau\tau}^{\rm B1}&=&-A_1\frac{c_{23}B_2+ s_{23}C_2}{c_{23}B_1 + s_{23}C_1} + A_2, \nonumber 
\label{Eq:f_a_B1}
\end{eqnarray}
and
\begin{eqnarray}
f_1^{\rm B1}  &=& A_1\frac{ \frac{t_{12}c_{23}}{c_{13}} +t_{13}s_{23}e^{i\delta} }{c_{23}B_1 + s_{23}C_1} + 1,
\nonumber \\
f_2^{\rm B1}  &=&  A_1\frac{ -\frac{c_{23}}{c_{13}t_{12}} +t_{13}s_{23}e^{i\delta} }{c_{23}B_1 + s_{23}C_1} + 1,
\nonumber \\
f_3^{\rm B1}  &=& A_1\frac{ -\frac{s_{23}}{t_{13}}e^{-i\delta} }{c_{23}B_1 + s_{23}C_1} + e^{-2i\delta},
\label{Eq:f1_f2_f3_B1}
\end{eqnarray}
as well as
\begin{eqnarray}
A_1&=&c_{23}^2+s_{23}^2 e^{-2i\delta}, \nonumber \\
A_2&=&s_{23}^2+c_{23}^2 e^{-2i\delta}, \nonumber \\
B_1&=& \frac{2c_{23}^2}{c_{13}\tan 2\theta_{12}} - t_{13}\sin 2\theta_{23} e^{-i\delta}, \nonumber \\
B_2&=& \frac{2s_{23}^2}{c_{13}\tan 2\theta_{12}} + t_{13}\sin 2\theta_{23} e^{-i\delta}, \nonumber \\
B_3&=&  \frac{\sin 2\theta_{23}}{c_{13}\tan 2\theta_{12}} +t_{13} \cos 2\theta_{23} e^{-i\delta}, \nonumber \\
C_1&=& 2  \left( \frac{s_{23}^2 e^{-i\delta}}{\tan 2\theta_{13}} -\frac{t_{13} c_{23}^2 e^{i\delta}}{2} \right), \nonumber \\
C_2&=& 2  \left( \frac{c_{23}^2 e^{-i\delta}}{\tan 2\theta_{13}} -\frac{t_{13} s_{23}^2 e^{i\delta}}{2} \right), \nonumber \\
C_3&=&  \sin 2\theta_{23} \left( \frac{e^{-i\delta}}{\tan 2\theta_{13}} +\frac{t_{13} e^{i\delta}}{2} \right),
\end{eqnarray}
with $c_{ij}=\cos\theta_{ij}$, $s_{ij}=\sin\theta_{ij}$ and $t_{ij}=\tan\theta_{ij}$. 

Similarly, we obtain
\begin{eqnarray}
f_{e\mu}^{\rm B2} &=& f_{\tau\tau}^{\rm B2}=0, 
\\
f_{e\tau}^{\rm B2}&=&\frac{A_2}{s_{23}B_2 - c_{23}C_2},
\nonumber \\
f_{\mu\mu}^{\rm B2}&=&A_2\frac{-s_{23}B_1+ c_{23}C_1}{s_{23}B_2 - c_{23}C_2} + A_1,
\nonumber \\
f_{\mu\tau}^{\rm B2}&=&A_2\frac{s_{23}B_3 + c_{23}C_3 }{s_{23}B_2 - c_{23}C_2} - \frac{1-e^{-2i\delta}}{2} \sin 2\theta_{23},
\nonumber 
\label{Eq:f_a_B2}
\end{eqnarray}
and
\begin{eqnarray}
f_1^{\rm B2}  &=&A_2\frac{ \frac{t_{12}s_{23}}{c_{13}} -t_{13}c_{23}e^{i\delta} }{s_{23}B_2 - c_{23}C_2} + 1,
\nonumber \\
f_2^{\rm B2} &=&A_2\frac{ -\frac{s_{23}}{c_{13}t_{12}} -t_{13}c_{23}e^{i\delta} }{s_{23}B_2 - c_{23}C_2} + 1,
\nonumber \\
f_3^{\rm B2} &=&A_2\frac{ \frac{c_{23}}{t_{13}}e^{-i\delta} }{s_{23}B_2 - c_{23}C_2} + e^{-2i\delta},
\label{Eq:m_B2}
\end{eqnarray}
for the ${\rm B_2}$ texture, 
\begin{eqnarray}
f_{e\mu}^{\rm B3} &=&f_{\mu\mu}^{\rm B1}=0, 
\\
f_{e\tau}^{\rm B3}&=&\frac{A_1}{s_{23}B_1 - c_{23}C_1},
\nonumber \\
f_{\mu\tau}^{\rm B3} &=&A_1\frac{s_{23}B_3 + c_{23}C_3 }{s_{23}B_1 - c_{23}C_1} - \frac{1-e^{-2i\delta}}{2} \sin 2\theta_{23},
\nonumber \\
f_{\tau\tau}^{\rm B3}&=&2A_1\frac{(s_{23}B_3+ c_{23}C_3)\cos 2\theta_{23} - s_{23}t_{13}e^{-i\delta}}{\sin 2\theta_{23}(s_{23}B_1 - c_{23}C_1)} - \frac{1-e^{-2i\delta}}{2} \cos 2\theta_{23}, \nonumber 
\label{Eq:f_a_B3}
\end{eqnarray}
and
\begin{eqnarray}
f_1^{\rm B3} &=& A_1\frac{ \frac{t_{12}s_{23}}{c_{13}} -t_{13}c_{23}e^{i\delta} }{s_{23}B_1 - c_{23}C_1} + 1,
\nonumber \\
f_2^{\rm B3}&=& A_1\frac{ -\frac{s_{23}}{c_{13}t_{12}} -t_{13}c_{23}e^{i\delta} }{s_{23}B_1 - c_{23}C_1} + 1,
\nonumber \\
f_3^{\rm B3}&=& A_1\frac{ \frac{c_{23}}{t_{13}}e^{-i\delta} }{s_{23}B_1 - c_{23}C_1} + e^{-2i\delta},
\label{Eq:m_B3}
\end{eqnarray}
for the ${\rm B_3}$ texture and
\begin{eqnarray}
f_{e\tau}^{\rm B4} &=&f_{\tau\tau}^{\rm B4}=0, 
\\
f_{e\mu}^{\rm B4}&=&-\frac{A_2}{c_{23}B_2 + s_{23}C_2},
\nonumber \\
f_{\mu\mu}^{\rm B4}&=&-A_2\frac{c_{23}B_1+ s_{23}C_1}{c_{23}B_2 + s_{23}C_2} + A_1,
\nonumber \\
f_{\mu\tau}^{\rm B4}&=&-A_2\frac{-c_{23}B_3 + s_{23}C_3 }{c_{23}B_2 + s_{23}C_2} - \frac{1-e^{-2i\delta}}{2} \sin 2\theta_{23}, \nonumber
\label{Eq:f_a_B4}
\end{eqnarray}
and
\begin{eqnarray}
f_1^{\rm B4}&=&A_2\frac{ \frac{t_{12}c_{23}}{c_{13}} + t_{13}s_{23}e^{i\delta} }{c_{23}B_2 + s_{23}C_2} + 1,
\nonumber \\
f_2^{\rm B4}&=&-A_2\frac{ \frac{c_{23}}{c_{13}t_{12}} -t_{13}s_{23}e^{i\delta} }{c_{23}B_2 + s_{23}C_2} + 1,
\nonumber \\
f_3^{\rm B4}&=&-A_2\frac{ \frac{s_{23}}{t_{13}}e^{-i\delta} }{c_{23}B_2 + s_{23}C_2} + e^{-2i\delta},
\label{Eq:m_B4}
\end{eqnarray}
for the ${\rm B_4}$ texture.

\section{\label{sec:Dark_matter} Radiative seesaw dark matter and neutrinoless double beta decay}

\subsection{$M_{ee}$ and dark matter}
In order to study the relation of the neutrino flavor mass $M_{\alpha\beta}$ and dark matter mass $M_1$, the Yukawa couplings
\begin{eqnarray}
{\bf h}=\left( 
\begin{array}{ccc}
h_{e1} & h_{e2} & h_{e3} \\
h_{\mu 1} & h_{\mu 2} & h_{\mu 3} \\
h_{\tau 1} & h_{\tau 2} & h_{\tau 3} \\
\end{array}
\right),
\end{eqnarray}
should be determined. For examples, the texture of the Yukawa matrix ${\bf h}$ is assumed as 
\begin{eqnarray}
{\bf h}=\left( 
\begin{array}{ccc}
0 & 0 & h_{e3} \\
h_{\mu 1} & h_{\mu 2} & h_{e 3} \\
h_{\mu 1} & h_{\mu 2} & -h_{e 3} \\
\end{array}
\right),
\label{Eq:h_Suematsu}
\end{eqnarray}
in Ref.\cite{Suematsu2010PRD} or parametrized with three parameters $h_1,h_2,h_3$ as
\begin{eqnarray}
{\bf h}=\left( 
\begin{array}{ccc}
h_1 & h_2 & h_3 \\
-0.68h_1 & h_2 & 3.56h_3 \\
0.31h_1 & -h_2 & 4.55h_3 \\
\end{array}
\right),
\label{Eq:h_Singirala}
\end{eqnarray}
in Ref.\cite{Singirala2017CPC}．

In this paper, we focus our attention on the relation between $M_{ee}$ and $M_1$.  For our purpose, assuming two zero elements in the flavor neutrino mass matrix, e.g. two zero texture, is one of the useful assumptions. The clear dependence of the flavor neutrino mass $M_{\alpha\beta}$ on $M_{ee}$ is obtained as Eq.(\ref{Eq:M_flavor}) and we can use not only
\begin{eqnarray}
M_{ee} = \sum_{k=1}^3 (h_{e k}^{\rm X})^2 \Lambda_k,
\end{eqnarray}
but also
\begin{eqnarray}
M_{ee} &=& (f_{\alpha\beta}^{\rm X})^{-1} M_{\alpha\beta} \nonumber \\
&=& (f_{\alpha\beta}^{\rm X})^{-1} \sum_{k=1}^3 h_{\alpha k}^{\rm X} h_{\beta k}^{\rm X} \Lambda_k,
\label{Eq:Mee_Lambda}
\end{eqnarray}
to reveal the the relation between $M_{ee}$ and $M_1$ (recall that $\Lambda_k = f(M_1)$). 

The derivation of the relation in Eq.(\ref{Eq:Mee_Lambda}) will be more difficult if we use the general parametrization of neutrino mixing, e.g., the Casas-Espinosa-Ibarra-Navarro parametrization \cite{CasasIbarra1999NPB,CasasIbarra2001NPB}. Thanks to our formula of Eq.(\ref{Eq:M_flavor}) for two zero texture, we find the clear linkage between dark matter mass $M_1$ in the one-loop radiative seesaw model and the effective neutrino mass for the neutrino less double beta decay $M_{ee}$ in Eq.(\ref{Eq:Mee_Lambda}). This result has been already numerically suggested by Kubo et. al. \cite{one_loop_Kubo2006PLB}. We confirm their suggestion by deriving analytical expressions for the two zero flavor neutrino mass texture. This is the main advantage of this paper.

For example, for $B_1$ case, the relations
\begin{eqnarray}
&& M_{e\tau} = 0,\quad M_{\mu\mu} = 0, \\
&& M_{ee} = (f_{e\mu}^{\rm B1})^{-1} M_{e\mu} = (f_{\mu\tau}^{\rm B1})^{-1}M_{\mu\tau} = (f_{\tau\tau}^{\rm B1})^{-1}M_{\tau\tau}, 
\nonumber
\end{eqnarray}
yield the following non-linear simultaneous equations 
\begin{eqnarray}
0&=&\sum_{k=1}^3 h_{e k}^{\rm B1} h_{\tau k}^{\rm B1} \Lambda_k=\sum_{k=1}^3 (h_{\mu k}^{\rm B1})^2 \Lambda_k,\nonumber \\
M_{ee} &=& \sum_{k=1}^3 (h_{e k}^{\rm B1})^2 \Lambda_k \nonumber \\
&=& (f_{e\mu}^{\rm B1})^{-1}\sum_{k=1}^3 h_{e k}^{\rm B1} h_{\mu k}^{\rm B1}\Lambda_k  \nonumber \\
&=& (f_{\mu\tau}^{\rm B1})^{-1}\sum_{k=1}^3 h_{\mu k}^{\rm B1} h_{\tau k}^{\rm B1}\Lambda_k  \nonumber \\
&=& (f_{\tau\tau}^{\rm B1})^{-1}\sum_{k=1}^3 (h_{\tau k}^{\rm B1})^2 \Lambda_k.
\label{Eq:nonlinear}
\end{eqnarray}
We can use the coupled equations in Eq.(\ref{Eq:nonlinear}) to reduce the number of assumptions for the Yukawa couplings.

\subsection{A specific case}
First, we take an extremely specific configuration of the Yukawa couplings to show a clear visible linkage between neutrinoless double beta decay and the mass of the dark matter candidate. More general analysis will be shown in the next subsection.

For the sake of simplicity, we assume that $N_{1,2,3}$ are nearly degenerate and we take $M_1 \sim M_2 \sim M_3 \sim M_0$ \cite{one_loop_Kubo2006PLB}.  In this case, we obtain the relation of $\Lambda_1=\Lambda_2=\Lambda_3$ and
\begin{eqnarray}
\tilde{M}_{ee} = (f_{\alpha\beta}^{\rm X})^{-1} \sum_{k=1}^3 h_{\alpha k}^{\rm X} h_{\beta k}^{\rm X},
\end{eqnarray}
from (\ref{Eq:Mee_Lambda}), where
\begin{eqnarray}
\tilde{M}_{ee} = \frac{8\pi^2}{\lambda_5 v^2}\frac{m^2_0-M^2_0}{M_0}\left(1-\frac{M^2_0}{m^2_0-M^2_0}\ln\frac{m_0^2}{M^2_0} \right)^{-1} M_{ee}.
\end{eqnarray}
One can readily find the following clear linkage between dark matter mass $M_0$ in the one-loop radiative seesaw model and the effective neutrino mass for the neutrino less double beta decay $M_{ee}$, which is described by
\begin{eqnarray}
M_{ee} &=& \frac{\sum_k h_{\alpha k}^{\rm X} h_{\beta k}^{\rm X}}{f_{\alpha\beta}^{\rm X}} 
 \frac{\lambda_5 v^2}{8\pi^2}\frac{M_0}{m^2_0-M^2_0} \left(1-\frac{M^2_0}{m^2_0-M^2_0}\ln\frac{m_0^2}{M^2_0} \right).
\label{Eq:Mee_M0}
\end{eqnarray}

As more specific example, we consider the possibility that two Yukawa couplings vanish. In the ${\rm B_1}$ case, we obtain the following non-linear simultaneous equations
\begin{eqnarray}
\label{Eq:nonlinear_B1}
&& \sum_{k=1}^3 (h_{e k}^{\rm B1})^2 = \tilde{M}_{ee}, \quad
\sum_{k=1}^3 h_{e k}^{\rm B1}h_{\mu k}^{\rm B1}  = f_{e\mu}^{B1} \tilde{M}_{ee}, \nonumber \\
&& \sum_{k=1}^3 h_{e k}^{\rm B1}h_{\tau k}^{\rm B1}  = 0, \quad
\sum_{k=1}^3( h_{\mu k}^{\rm B1})^2  = 0, \\
&& \sum_{k=1}^3 h_{\mu k}^{\rm B1}h_{\tau k}^{\rm B1} = f_{\mu\tau}^{B1} \tilde{M}_{ee}, \quad
\sum_{k=1}^3 (h_{\tau k}^{\rm B1})^2  = f_{\tau\tau}^{B1} \tilde{M}_{ee}.
\nonumber 
\end{eqnarray}
A solution to the non-linear simultaneous equations in Eq.(\ref{Eq:nonlinear_B1}) is obtained as
\begin{eqnarray}
h_{e1}^{\rm B1}&=&\frac{ (f_{e\mu}^{\rm B1})^2 \tilde{M}_{ee} + (h_{\mu 1}^{\rm B1})^2 \left(\frac{(f_{\mu\tau}^{\rm B1})^2}{(f_{\mu\tau}^{\rm B1})^2 \tilde{M}_{ee} - f_{\tau\tau}^{\rm B1} (h_{\mu 1}^{B1})^2} +1\right) }{2f_{e\mu}^{\rm B1} h_{\mu 1}^{B1}}, \nonumber \\
h_{e2}^{\rm B1}&=&-\frac{f_{\mu\tau}^{\rm B1} \tilde{M}_{ee} }{h_{\mu 1}^{B1} \sqrt[]{\mathstrut \tilde{M}_{ee} (f_{\tau\tau}^{\rm B1}-\frac{(f_{\mu\tau}^{\rm B1})^2 \tilde{M}_{ee}}{(h_{\mu 1}^{B1})^2}} )}, \nonumber \\
h_{e3}^{\rm B1}&=&-\frac{i}{2f_{e\mu}^{\rm B1} h_{\mu 1}^{B1}} \left((f_{e\mu}^{\rm B1})^2 \tilde{M}_{ee} \right. \nonumber \\
&& \left. + (h_{\mu 1}^{\rm B1})^2 \left(\frac{(f_{\mu\tau}^{\rm B1})^2}{f_{\tau\tau}^{\rm B1} (h_{\mu 1}^{B1})^2 - (f_{\mu\tau}^{\rm B1})^2 \tilde{M}_{ee}} -1\right) \right), \nonumber \\
h_{\mu 1}^{\rm B1}&:& {\rm free \ parameter}, \quad
h_{\mu 2}^{\rm B1}= 0, \quad
h_{\mu 3}^{\rm B1}= ih_{\mu 1}^{\rm B1}, \nonumber \\
h_{\tau 1}^{\rm B1}&=&\frac{f_{\mu\tau}^{\rm B1} \tilde{M}_{ee}}{h_{\mu 1}^{B1}}, \quad
h_{\tau 2}^{\rm B1}=\sqrt[]{\mathstrut \tilde{M}_{ee} (f_{\tau\tau}^{\rm B1}-\frac{(f_{\mu\tau}^{\rm B1})^2 \tilde{M}_{ee}}{(h_{\mu 1}^{B1})^2}} ), \nonumber \\
h_{\tau 3}^{\rm B1}&=&0,
\label{Eq:specific_h_B1}
\end{eqnarray}
where $h_{\mu 1}^{\rm B1}$ remains as a free parameter. Similarly, we can obtain solutions for the ${\rm B_2}$, ${\rm B_3}$ and ${\rm B_4}$ cases.

We note that there are only three direct assumptions for the Yukawa couplings: (1) $h_{\mu 2}^{\rm B1}=0$, (2) $h_{\tau 3}^{\rm B1}=0$ and (3) $h_{\mu 1}^{\rm B1}$ as a free parameter. This configuration is specific, but there are less assumptions for the Yukawa couplings in Eq.(\ref{Eq:specific_h_B1}) compared with these in Eq.(\ref{Eq:h_Suematsu}) or Eq.(\ref{Eq:h_Singirala}). We use the coupled equations in Eq.(\ref{Eq:nonlinear}) or Eq.(\ref{Eq:nonlinear_B1}) instead of some direct assumptions for the Yukawa couplings. 

Although, we have reached our main goal of this paper to analytically show the linkage between the dark matter mass and the effective neutrino mass for the two zero flavor neutrino mass texture as in Eq. (\ref{Eq:Mee_Lambda}), some additional numerical calculations may be required to visually confirm the validity of our method. 

In the neutrino sector, we use the following global fit of mixing angles \cite{Esteban2017JHEP}
\begin{eqnarray}
\sin^2 \theta_{12} &=& 0.306,
\nonumber\\
\sin^2 \theta_{23} &=& 0.441 \ {\rm or} \ 0.587,
\nonumber\\
\sin^2 \theta_{13} &=& 0.0217.
\label{Eq:observedangle}
\end{eqnarray}
Although the CP-violating Dirac phase is reported to be \cite{Esteban2017JHEP}:
\begin{eqnarray}
\delta /^\circ &=& 261^{+51}_{-59} \ ({\rm  NO}) \ {\rm or} \ 277^{+40}_{-46} \ ({\rm  IO}),
\label{Eq:observedDiracPhase}
\end{eqnarray}
we vary $\delta$ as $0^\circ \le \delta \le 360^\circ$ in this subsection. The estimated upper limit of the magnitude of the effective neutrino mass from the experiments is $\vert M_{ee} \vert \le 0.20-2.5$ eV \cite{Cremonesi2014AHEP}, we assume $0.01$ eV $\le M_{ee} \le 1$ eV. Recently, NOvA collaboration has reported that the hypothesis of the inverted mass hierarchy with $\theta_{23}$ in the lower octant is disfavored at greater than $93\%$ C.L. for all values of $\delta$ \cite{NOvA2017PRL}; however, the problem of the octant of $\theta_{23}$ (i.e. lower octant $\theta_{23}<45^\circ$ or upper octant $\theta_{23}>45^\circ$) is still unresolved. According to Dev et al. \cite{Dev2014PRD} and Meloni et al. \cite{Meloni2014PRD}, we employ the values of $\theta_{23}$ given for both octant which are shown in TABLE \ref{table1}. 

In the dark sector, we adopt the following standard criteria \cite{one_loop_Kubo2006PLB,Ibarra2016PRD}: (1) The quartic coupling satisfy the relation of $|\lambda_5| \ll 1$ for small neutrino masses. (2) The Yukawa couplings are sizeable. (3) The masses of new fields lie in the range between a few GeV and a few TeV.  Since we assume that the additional Majorana fermion is dark matter, we require the relation of $M_0 < m_0$. We take 
\begin{eqnarray}
&& 10^{-10} \le \lambda_5 \le 10^{-6}, \nonumber \\
&& 0.01 \le h_{xx}^{\rm X} \le 1.0, \nonumber \\
&& 100 {\rm GeV} \le M_1 \le 10 {\rm TeV}, \nonumber \\
&& m_0=1.5M_0,
\label{Eq:DMparameteter}
\end{eqnarray}

We estimate the allowed dark matter mass $M_0$ and effective neutrino mass $M_{ee}$ for $\Omega h^2 = 0.1184 \pm 0.0012$ \cite{Planck2016AA} and $ {\rm Br}(\mu \rightarrow e\gamma) \le 4.2 \times 10^{-13}$ \cite{MEG2016arXiv}. Although the upper limits of the branching ratio of ${\rm Br}(\tau \rightarrow \mu \gamma) \le 4.4 \times 10^{-8}$ and ${\rm Br}(\tau \rightarrow e \gamma) \le 3.3 \times 10^{-8}$ are also reported \cite{BABAR2010PRL}, we only account for ${\rm Br}(\mu \rightarrow e\gamma)$ since it is the most stringent constraint. Since the flavor mixing effect is small, we assume that the charged lepton mass matrix is diagonal.  

\begin{figure}[t]
\begin{center}
\includegraphics[width=8.0cm]{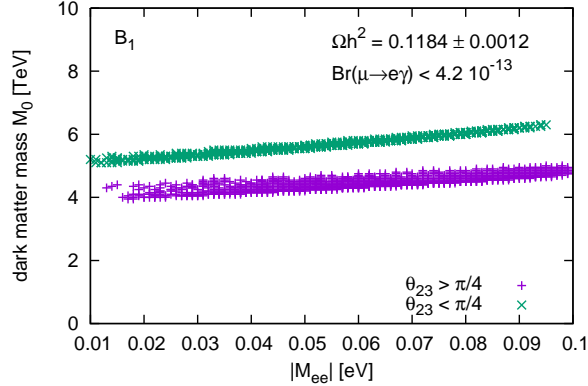}
\caption{Dark matter mass $M_0$ v.s. effective neutrino mass $M_{ee}$ for $\Omega h^2 = 0.1184 \pm 0.0012$ and $ {\rm Br}(\mu \rightarrow e\gamma) \le 4.2 \times 10^{-13}$ in the case of ${\rm B_1}$ texture with $h_{\mu 2}^{\rm B1}=0$ and $h_{\tau 3}^{\rm B1}=0$.}
\label{figure1}
\end{center}
\end{figure}

FIG.\ref{figure1} shows the allowed dark matter mass $M_0$ and effective neutrino mass $M_{ee}$ for $\Omega h^2 = 0.1184 \pm 0.0012$  and $ {\rm Br}(\mu \rightarrow e\gamma) \le 4.2 \times 10^{-13}$ with three direct assumptions for the Yukawa couplings: (1) $h_{\mu 2}^{\rm B1}=0$, (2) $h_{\tau 3}^{\rm B1}=0$ and (3) $h_{\mu 1}^{\rm B1}$ as a free parameter. The purpose of this paper is to show a linkage between neutrinoless double beta decay and the mass of the dark matter candidate encoded in Eq. (\ref{Eq:Mee_Lambda}). In FIG.\ref{figure1}, there are visible correlations between dark matter mass $M_0$ and the effective neutrino mass of the neutrinoless double beta decay $M_{ee}$. As we will address later, if we include the vacuum stability bound to our analysis, the dark matter mass should be below $\sim 1$ TeV. In this subsection, we take an extremely specific and simple configuration of the Yukawa couplings to just illustrate the visible correlations between $M_0$ and $M_{ee}$.

\begin{table}[t]
\tbl{Octant of $\theta_{23}$. }
{\begin{tabular}{c|cccc}
\hline
NO/IO     & $B_1$ & $B_2$ & $B_3$ & $B_4$ \\
\hline
NO & $\ell$ & $u$ & $\ell$ & $u$ \\
IO & $u$ & $\ell$ & $u$ & $\ell$ \\
\hline
\end{tabular}
\label{table1}}
\end{table}


\subsection{\label{subsec:More_General_case}More general analysis}
Next, we preform the numerical analysis with more general setup. In this subsection, the assumption of ``two Yukawa couplings vanish" in the previous subsection is removed.  All Yukawa couplings are determined by numerical calculations of non-linear simultaneous equations in Eq.(\ref{Eq:nonlinear}) by Monte Calro method. In this subsection, our assumptions for the values of parameters in the neutrino sector and the dark sector are the same of those in the previous subsection except the following mass relation
\begin{eqnarray}
 M_1 < M_3 \sim m_0,
\label{Eq:DMparameteter}
\end{eqnarray}
instead of $m_0=1.5M_0$ and perform the numerical analysis for the following two cases: (i) $M_1 \ll M_2 < M_3$ (at least $M_2 \ge 20 M_1$) and (ii) $M_1 \lesssim M_2 < M_3$.

\begin{figure}[t]
\begin{center}
\includegraphics[width=8.0cm]{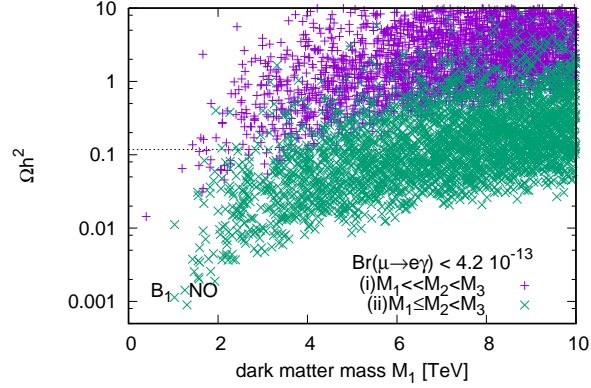}
\caption{Relic abundance of Dark matter  $\Omega h^2$ v.s. dark matter mass $M_1$  for ${\rm B_1}$ texture in the case of NO.}
\label{fig:OmegaM1}
\end{center}
\end{figure}
\begin{figure}[t]
\begin{center}
\includegraphics[width=8.0cm]{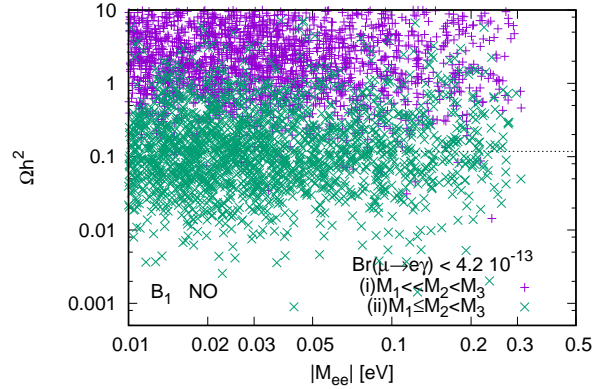}
\caption{Same as FIG.\ref{fig:OmegaM1} but relic abundance of Dark matter $\Omega h^2$ v.s.  effective neutrino mass $M_{ee}$.}
\label{fig:OmegaMee}
\end{center}
\end{figure}
\begin{figure}[t]
\begin{center}
\includegraphics[width=8.0cm]{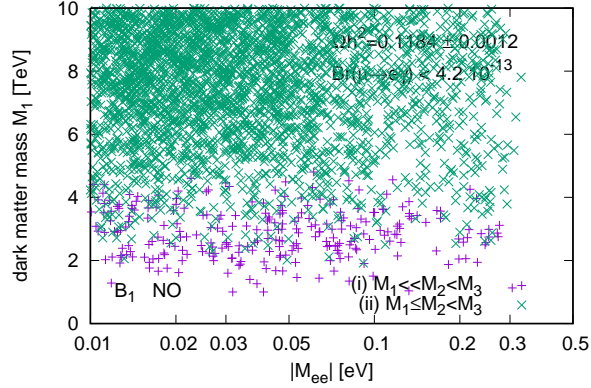}
\caption{Dark matter mass $M_1$ v.s. effective neutrino mass $M_{ee}$ for ${\rm B_1}$ texture.}
\label{fig:M1Mee}
\end{center}
\end{figure}

\begin{table}[t]
\tbl{Arrowed region of Dark matter mass $M_1$ and upper bound of effective neutrino mass $M_{ee}$ for $\Omega h^2 = 0.1184 \pm 0.0012$ \cite{Planck2016AA}. }
{\begin{tabular}{cccccc}
\hline
${\rm B_X}$ & NO/IO & $M_1$,$M_2$ & $M_1$ [TeV] & $|M_{ee}|_{\rm max}$ [eV] \\
\hline
${\rm B_1}$ & NO & $M_1 \lesssim M_2$ & 2.09-10.0 & 0.274\\
      & NO& $M_1 \ll M_2$  & 1.08-4.36 & 0.318\\
      & IO& $M_1 \lesssim M_2$  & 1.65-9.99 & 0.270\\
      & IO& $M_1 \ll M_2$  & 1.26-4.67 & 0.221\\
\hline
${\rm B_2}$ & NO & $M_1 \lesssim M_2$ & 3.10-9.94 & 0.121\\
      & NO& $M_1 \ll M_2$  & 1.16-4.92 & 0.297\\
      & IO& $M_1 \lesssim M_2$  & 2.52-9.99 & 0.312\\
      & IO& $M_1 \ll M_2$  & 1.21-4.81 & 0.312\\
\hline
${\rm B_3}$ & NO & $M_1 \lesssim M_2$ & 2.57-9.98 & 0.308\\
      & NO& $M_1 \ll M_2$  & 1.00-4.89 & 0.235\\
      & IO& $M_1 \lesssim M_2$  & 2.53-10.0 & 0.246\\
      & IO& $M_1 \ll M_2$  & 1.15-4.75 & 0.292\\
\hline
${\rm B_4}$ & NO & $M_1 \lesssim M_2$ & 2.54-9.98 & 0.233\\
      & NO& $M_1 \ll M_2$  & 1.09-4.72 & 0.307\\
      & IO& $M_1 \lesssim M_2$  & 1.73-9.98 & 0.282\\
      & IO& $M_1 \ll M_2$  & 1.16-4.70 & 0.308\\
\hline
\end{tabular}
\label{table2}}
\end{table}

FIG. \ref{fig:OmegaM1} shows the relic abundance of Dark matter  $\Omega h^2$ v.s. dark matter mass $M_1$ for ${\rm B_1}$ texture in the case of NO for $ {\rm Br}(\mu \rightarrow e\gamma) \le 4.2 \times 10^{-13}$ \cite{MEG2016arXiv}.  The horizontal dotted line in the figures show the observed relic abundance $\Omega h^2 \sim 0.1$. FIG. \ref{fig:OmegaMee} shows the same as FIG.\ref{fig:OmegaM1} but relic abundance of Dark matter  $\Omega h^2$ v.s.  effective neutrino mass $M_{ee}$. FIG. \ref{fig:M1Mee} shows the Dark matter mass $M_1$ v.s. effective neutrino mass $M_{ee}$ (we plot in the range of $0.01$ eV $\le M_{ee} \le 0.1$ eV) for ${\rm B_1}$ texture for $\Omega h^2 = 0.1184 \pm 0.0012$ and $ {\rm Br}(\mu \rightarrow e\gamma) \le 4.2 \times 10^{-13}$. TABLE \ref{table2} shows the arrowed region of Dark matter mass $M_1$ and upper bound of effective neutrino mass $M_{ee}$ for $\Omega h^2 = 0.1184 \pm 0.0012$ \cite{Planck2016AA} and $ {\rm Br}(\mu \rightarrow e\gamma) \le 4.2 \times 10^{-13}$ \cite{MEG2016arXiv} in all case of ${\rm B_1, B_2,B_3}$ and ${\rm B_4}$.

From FIG. \ref{fig:OmegaM1}, FIG. \ref{fig:OmegaMee}, FIG. \ref{fig:M1Mee} and TABLE \ref{table2}, we find the following:
\begin{itemize}
\item The dark matter with $M_1 \sim \mathcal{O}({\rm TeV})$ is consistent with the relic abundance of dark matter. This is commonly expected feature of the heavy cold dark matter models \cite{DMreviews_Baer2015PREP,DMreviews_Arcadi2017arXiv}.
\item More relic abundance of dark matter $\Omega h^2$ is obtained in the case (i) compared with it in the case (ii). In the case (ii), coannihilation channel is allowed because of $M_1 \sim M_2$ and this coannihilation channel yields the reduction of relic abundance. 
\item $|M_{ee}|<\mathcal{O}(0.1)$eV. In next subsection, we show that $|M_{ee}| \sim \mathcal{O}(0.01)$ eV is expected. 
\item There is no significant distinction for $M_1$ and $|M_{ee}|$ in NO and IO for all textures ${\rm B_1,B_2,B_3}$ and ${\rm B_4}$. Since the nearly degenerate neutrino mass pattern is obtained in the texture two zeros \cite{Dev2014PRD,KitabayashiYasue2016IJMPA,KitabayashiYasue2016PRD} and we take wide parameter range for dark side as in Eq.(\ref{Eq:DMparameteter}), the difference between NO and IO for $M_1$ and $M_{ee}$ to be small.
\item While there are algebraic relations neutrinoless double beta decay and the mass of the dark matter candidate encoded in Eq. (\ref{Eq:Mee_Lambda}) for all textures ${\rm B_1,B_2,B_3}$ and ${\rm B_4}$, because of our general parameter setup, these relation are washed out by other parameters. Recall that, for more simple parameter setup, see FIG. \ref{figure1}, there are visible correlations between dark matter mass $M_0$ and the effective neutrino mass of the neutrinoless double beta decay $M_{ee}$.
\item There is no significant distinction for $M_1$ and $|M_{ee}|$ in textures ${\rm B_1,B_2,B_3}$ and ${\rm B_4}$. We comment that according to Ref.\cite{Zhou2016CPC}, ${\rm B_2}$ and ${\rm B_4}$ textures are compatible with the recent neutrino oscillation data for $\theta_{23}>45^\circ$ and $|M_{ee}| \neq 0$ ; however, the problem of the octant of $\theta_{23}$ is still unresolved. Thus we estimate $M_1$ and $|M_{ee}|$ for all textures ${\rm B_1,B_2,B_3}$ and ${\rm B_4}$.
\item We would like to comment that in the paper by Lindner et al \cite{Lindner2016PRD} the vacuum stability of the scotogenic model is discussed and they find hardly any viable points for a dark matter mass above 1.2 TeV with their parameter setup. These vacuum stability bound is very important and relevant for our study and we should include the stability bound. If we take account these stability bound, the large dark matter masses shown in TABLE \ref{table2} will be suppressed and the allowed dark matter mass should be only around 1 TeV. The detailed analysis will be appeared in our future study.
\end{itemize}
%

\subsection{\label{subsec:Majorana}CP violating phases}
Finally, we study the dependence of the CP-violating Majorana phases $\alpha_2$ and $\alpha_3$ on the CP-violating Dirac phase $\delta$. 

Since the mass eigenvalue $m_j$ is obtained as Eq.(\ref{Eq:mass_ev}), the Majorana phase $\phi_j$ depends on not only $\delta$ but also ${\rm arg} (M_{ee})$ as follows \cite{KitabayashiYasue2016IJMPA,KitabayashiYasue2016PRD};
\begin{eqnarray}
\phi_j = -{\rm arg}(m_j e^{-i \phi_j}) = -{\rm arg} (f_j^{\rm X}(\delta)) - {\rm arg} (M_{ee}).
\end{eqnarray}
On the contrary, the physical CP-violating Majorana phase $\alpha_2,\alpha_3$ is specified by two combinations made of $\phi_1,\phi_2,\phi_3$ such as  
\begin{eqnarray}
\alpha_j = \phi_j - \phi_1 = {\rm arg}\left( \frac{f_1^{\rm X}(\delta)}{f_j^{\rm X}(\delta)} \right),
\label{Eq:MajoranaPhases}
\end{eqnarray}
which depends on the CP-violating Dirac phase $\delta$ in all cases of $B_1,B_2,B_3$ and $B_4$. 

For example, we have the following the ratio of $f_1^{\rm X}(\delta)/f_j^{\rm X}(\delta)$ for the ${\rm B_1}$ texture
\begin{eqnarray}
\left. \frac{f_1^{\rm B1}}{f_2^{\rm B1}} \right|_{{\rm NO}} &\simeq& -2.27+\frac{1}{0.287-0.0452 e^{i\delta}} +\frac{1}{-0.602+2.13 e^{i\delta}},
\nonumber \\
\left. \frac{f_1^{\rm B1}}{f_3^{\rm B1}} \right|_{{\rm NO}} &\simeq& - 0.763-0.250 e^{i\delta} + \frac{1}{-0.669-5.61 e^{i\delta} + (0.505-1.95 e^{i\delta})^{-1} }, 
\end{eqnarray}
with the mixing angles in Eq.(\ref{Eq:observedangle}) in the case of NO and
\begin{eqnarray}
\left. \frac{f_1^{\rm B1}}{f_2^{\rm B1}} \right|_{{\rm IO}} &\simeq& -2.27+\frac{1}{0.301-0.0186 e^{i\delta}} +\frac{1}{-0.667+3.35 e^{i\delta}},
\nonumber \\
\left. \frac{f_1^{\rm B1}}{f_3^{\rm B1}} \right|_{{\rm IO}} &\simeq& - 1.37-0.186 e^{i\delta} + \frac{1}{-1.27-4.08 e^{i\delta} + (0.500-1.21 e^{i\delta})^{-1} }, 
\end{eqnarray}
in the case of IO.

Similarly, we obtain
\begin{eqnarray}
\left. \frac{f_1^{\rm B2}}{f_2^{\rm B2}} \right|_{{\rm NO}} &\simeq& -2.27+\frac{1}{-0.660-3.12 e^{i\delta}} +\frac{1}{0.300+0.0220 e^{i\delta}},
\nonumber \\
\left. \frac{f_1^{\rm B2}}{f_3^{\rm B2}} \right|_{{\rm NO}} &\simeq& -1.22+0.197 e^{i\delta} + \frac{1}{-1.15+4.34 e^{i\delta} + (0.508+1.36 e^{i\delta})^{-1} }, 
\end{eqnarray}
\begin{eqnarray}
\left. \frac{f_1^{\rm B2}}{f_2^{\rm B2}} \right|_{{\rm IO}} &\simeq& -2.27+\frac{1}{-0.577-1.88 e^{i\delta}} +\frac{1}{0.281+0.0539 e^{i\delta}},
\nonumber \\
\left. \frac{f_1^{\rm B2}}{f_3^{\rm B2}} \right|_{{\rm IO}} &\simeq& - 0.684-0.265 e^{i\delta} + \frac{1}{-0.546-5.95 e^{i\delta} + (0.500+2.08 e^{i\delta})^{-1} }, 
\end{eqnarray}
for the ${\rm B_2}$ texture
\begin{eqnarray}
\left. \frac{f_1^{\rm B3}}{f_2^{\rm B3}} \right|_{{\rm NO}} &\simeq& -2.26+\frac{1}{-0.647-3.29 e^{i\delta}} +\frac{1}{0.297+0.0326 e^{i\delta}},
\nonumber \\
\left. \frac{f_1^{\rm B3}}{f_3^{\rm B3}} \right|_{{\rm NO}} &\simeq& - 0.928+0.197 e^{i\delta} + \frac{1}{1.21+6.14 e^{i\delta}},
\end{eqnarray}
\begin{eqnarray}
\left. \frac{f_1^{\rm B3}}{f_2^{\rm B3}} \right|_{{\rm IO}} &\simeq& -2.26+\frac{1}{-0.65-2.44 e^{i\delta}} +\frac{1}{0.296+0.0244 e^{i\delta}},
\nonumber \\
\left. \frac{f_1^{\rm B3}}{f_3^{\rm B3}} \right|_{{\rm IO}} &\simeq& - 1.49+0.265 e^{i\delta} + \frac{1}{0.670+2.53 e^{i\delta}}, 
\end{eqnarray}
for the ${\rm B_3}$ texture and 
\begin{eqnarray}
\left. \frac{f_1^{\rm B4}}{f_2^{\rm B4}} \right|_{{\rm NO}} &\simeq& -2.27+\frac{1}{0.297-0.0257 e^{i\delta}} +\frac{1}{-0.647+2.59 e^{i\delta}},
\nonumber \\
\left. \frac{f_1^{\rm B4}}{f_3^{\rm B4}} \right|_{{\rm NO}} &\simeq& - 1.33-0.250 e^{i\delta} + \frac{1}{0.75-3.01e^{i\delta}},
 \end{eqnarray}
\begin{eqnarray}
\left. \frac{f_1^{\rm B4}}{f_2^{\rm B4}} \right|_{{\rm IO}} &\simeq& -2.27+\frac{1}{0.296-0.0346 e^{i\delta}} +\frac{1}{-0.647+3.47 e^{i\delta}},
\nonumber \\
\left. \frac{f_1^{\rm B4}}{f_3^{\rm B4}} \right|_{{\rm IO}} &\simeq& 0.738-0.186 e^{i\delta} + \frac{1}{1.35 - 7.26 e^{i\delta} }, 
\end{eqnarray}
for the ${\rm B_4}$ texture.

\begin{figure}[t]
\begin{center}
\includegraphics[width=8.0cm]{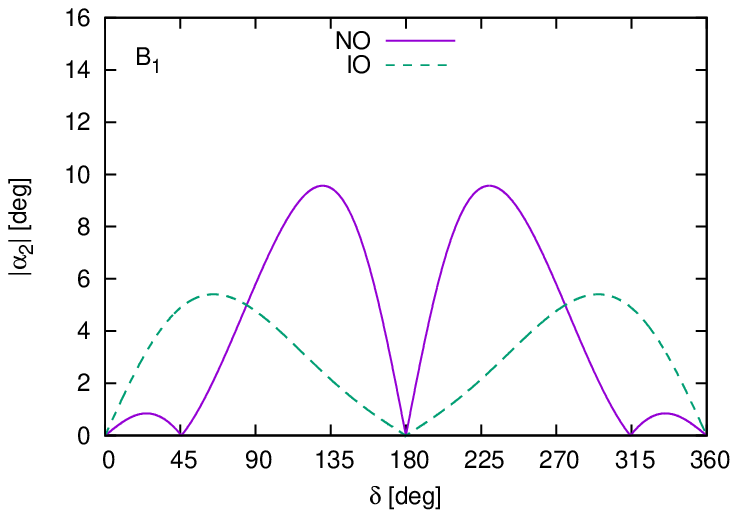}
\includegraphics[width=8.0cm]{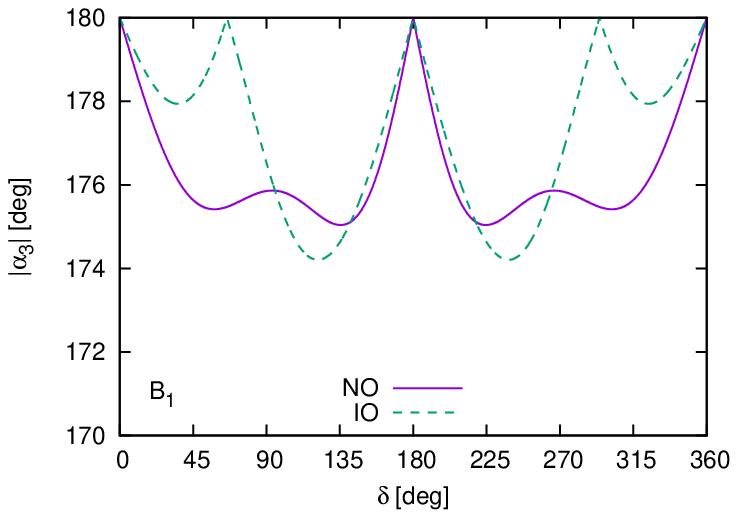}
\caption{The dependence of the Majorana CP-violating phases $\alpha_2$ (left panel) and $\alpha_3$ (right panel)on the CP-violating Dirac phase $\delta$ for ${\rm B_1}$ texture.  }
\label{fig:phase}
\end{center}
\end{figure}


\begin{table}[t]
\tbl{Majorana CP phases for $\delta = 261^\circ$ (NO) or $\delta = 277^\circ$ (IO). }
{\begin{tabular}{cccc}
\hline
${\rm B_X}$ & NO/IO & $|\alpha_2|/^\circ$ & $|\alpha_3|/^\circ$ \\
\hline
${\rm B_1}$ & NO & 7.07 & 176\\
      & IO & 5.06 & 178\\
\hline
${\rm B_2}$ & NO & 4.06 & 179\\
      & IO & 11.0 & 173\\
\hline
${\rm B_3}$ & NO & 3.48 & 178\\
      & IO & 5.66 & 176\\
\hline
${\rm B_4}$ & NO & 3.46 & 178\\
      & IO & 5.66 & 176\\
\hline
\end{tabular}
\label{table3}}
\end{table}

Shown in FIG.\ref{fig:phase} is the dependence of the Majorana CP-violating phases $\alpha_2$ (top panel) and $\alpha_3$ (bottom panel) on the CP-violating Dirac phase $\delta$ for ${\rm B_1}$ texture. If we fix the Dirac CP phase as
\begin{eqnarray}
\delta /^\circ &=& 261 \ ({\rm  NO}) \ {\rm or} \ 277 \ ({\rm  IO}),
\end{eqnarray}
the Majorana CP phases for ${\rm B_1,B_2,B_3}$ and ${\rm B_4}$ texture are obtained as in TABLE. \ref{table3}. 

To show the impact on $M_{ee}$ of the Dirac CP phase (the Majorana CP-violating phases), we reconsider the effective neutrino mass $M_{ee}$: 
\begin{eqnarray}
|M_{ee}|=|c_{12}^2c_{13}^2 m_1+s_{12}^2c_{13}^2m_2e^{i\alpha_2}+s_{13}^2m_3e^{i\alpha_3}|.
\end{eqnarray}
Since there is the relation of $\alpha_{2,3}=f(\delta)$, the effective neutrino mass is a function of the Dirac CP phase, e.g., $|M_{ee}|=f(\delta)$. For example, we take $(m_1,m_2,m_3)$ = $(6.27,6.33,7.97) \times 10^{-2}$ eV for NO and $(m_1,m_2,m_3)$ = $(6.68,6.73,4.61) \times 10^{-2}$ eV for IO in the case of ${\rm B1}$ texture \cite{Meloni2014PRD, KitabayashiYasue2016IJMPA,KitabayashiYasue2016PRD}.  With the CP-violating Majorana phases in TABLE. \ref{table3}, we obtain 
\begin{eqnarray}
|M_{ee}| \sim \begin{cases}
0.059\ {\rm eV} & {{\rm (NO)}}\\
0.064\ {\rm eV}& {{\rm (IO)}} \\
\end{cases}
\end{eqnarray}
in the ${\rm B_1}$ texture.

The upper limit of the magnitude of the effective neutrino mass is currently restricted to be $\vert M_{ee} \vert \le 0.20-2.5$ eV \cite{Cremonesi2014AHEP}; however, in the future experiments, a desired sensitivity $\vert M_{ee} \vert \simeq$ a few $10^{-2}$ eV will be reached \cite{Benato2015EPJC}. Upcoming experiments for neutrinoless double beta decay and dark matter search to be useful to probe the possible linkage between $M_{ee}$ and $M_1$. 

\section{\label{sec:summary}Summary}
We have shown the linkage between dark matter mass $M_1$ in the one-loop radiative seesaw model and the effective neutrino mass $M_{ee}$ for the neutrino less double beta decay for the two zero flavour neutrino mass texture. 

The neutrino flavor masses obtained for the one-loop radiative seesaw model (Eq.(\ref{Eq:M_alpha_beta})) and for two zero flavor neutrino mass texture (Eq.(\ref{Eq:M_flavor})). We have found that the clear linkage between dark matter mass in the one-loop radiative seesaw model and the effective neutrino mass for the neutrino less double beta decay (Eq.(\ref{Eq:Mee_Lambda})). This relation has been already numerically suggested \cite{one_loop_Kubo2006PLB}. We have confirmed this result more explicitly by deriving exact analytical expression. Supplemental numerical estimation of our analytical results is performed to visually confirm the validity of our method.



\end{document}